\documentclass[twocolumn,pre,showpacs]{revtex4-1}
\usepackage{graphicx}
\begin{document}
\title{Melting of genomic DNA: predictive modeling by nonlinear lattice dynamics}
%
%
\author{Nikos Theodorakopoulos$^{1,2}$ 
} 
\affiliation{
$^{1}$Theoretical and Physical Chemistry Institute, National Hellenic Research Foundation,\\
Vasileos Constantinou 48, 116 35 Athens, Greece \\
$^{2}$Fachbereich Physik, Universit\"at Konstanz, 78457 Konstanz, Germany
}
\date{\today}
\begin{abstract}
The melting behavior of long, heterogeneous  DNA chains is examined within the framework of the nonlinear lattice dynamics based Peyrard-Bishop-Dauxois (PBD) model. Data for the pBR322 plasmid and the complete T7 phage have been used to obtain model fits and determine parameter dependence on salt content. Melting curves predicted for the complete fd phage and the Y1 and Y2 fragments of the $\phi$X174 phage without any adjustable parameters are in good agreement with experiment. The calculated  probabilities for single base-pair opening are consistent with values obtained from imino proton exchange experiments.  
\end{abstract}
\maketitle
\section{Introduction}
The phenomenon of thermal  DNA denaturation can be interpreted in terms of mesoscopic, one-dimensional, statistical or nonlinear lattice-dynamical models. Models of the first type, based on the equilibrium statistical mechanics of the helix-coil transition, proposed and solved long ago by Poland  and Scheraga (PS) \cite{PoScheFi}
have been extensively studied, refined and developed by subsequent research; this substantial effort, incorporating a large body of enthalpic data, has led to algorithms for the detailed prediction of melting profiles with a rich internal structure \cite{WaBe85, Blake91}; as a result, the microscopic parametrization of this family of models is well advanced \cite{Blossey03}. Models of the second type, first proposed by Peyrard, Bishop and Dauxois (PBD)  \cite{PB, dauxpeyr1, dauxpeyr2}  were largely motivated by the study of dynamical properties, in particular those associated with the local breathing of DNA which characterizes the initiation of the transcription process; 
in the limiting case of infinite, homogeneous chains they describe DNA denaturation as an exact, one-dimensional thermodynamic phase transition \cite{DTPStatPhys}, whose effective (observable) order is controlled by the strength of the nonlinear base-stacking interaction \cite{CuleHwa,TDPprl} and can be followed in detail by finite-size scaling analysis \cite{helicoidal}. Curiously, although this type of model is uniquely suited \cite{CuleHwa, Zhang} to describe the sharp multistep melting process taking place at long sequences, where the effects of cooperativity are most likely to be correctly modeled with a small set of coupling parameters, comparisons of PBD-based DNA melting curves with experimental data have been confined to relatively short chains where multiphase melting is either absent \cite{CampaGian, Weber} or marginal \cite{PeyrCuesta, Montri, Ares2005}.  As a result, one of the major potential sources of experimental validation of the PBD model remains unexplored.

In this 
paper
I will present an analysis of melting profiles based on the standard version of the PBD model \cite{dauxpeyr1} for a number of specific sequences with many thousands of base pairs. I will use the parameters obtained by fitting two sets of published experimental data corresponding to different ionic contents,  the  pBR322 plasmid and the T7 phage melting profiles, to estimate the parameters' variation with ionic content. This procedure provides model parameters for any ionic content in the usual experimental range and thus paves the way for predicting the melting profile of any long DNA sequence. I present predicted melting profiles with no adjustable parameters and compare them with experimental data in three cases: the complete fd-phage, and the Y1 and Y2 fragments of the $\phi X174$ phage. The computed profiles, although not perfect, reflect accurately many of the actual complexities of sequence-dependent melting. 
Furthermore, the set of parameters obtained by fitting the melting profiles can be used to predict the probability of single base-pair opening at ambient temperatures. The calculated values, in the order of $1$ ppm, are in line with estimates obtained from imino proton exchange measurements \cite{Gueron87}. In summary, the PBD model seems to provide an excellent mesoscopic, dynamically motivated framework for the quantitative description of genomic DNA melting; this could be of particular use in further work where the model enjoys a comparative advantage, e.g. in extending the study of local DNA flexibility \cite{WeberFlex} to longer sequences.
 \section{Model notation and numerical procedure}
The configurational energy of a chain of $N$ base pairs is 
\begin{equation}
	H_P = \sum_{j=1}^{N-1} W(y_{j},y_{j+1})  + 
 \sum_{j=1}^{N} V_j(y_j) 
\label{eq:Ham}
\end{equation}
where the transverse coordinate $y_j$ represents the separation of the two bases at the $j$th site, the anharmonic elastic term 
\begin{equation}
	W(y_j,y_{j+1}) = \frac{k}{2}\left[ 1 + \rho e^{-b(y_j+y_{j+1})}  
\right](y_j-y_{j+1})^2
\label{eq:stack}
\end{equation}
models the nonlinear base-stacking interaction,
and the on-site Morse potential
$V_j(y_j) = D_j(1-e^{-\alpha_j y_j})^2$
describes the combined effects of hydrogen-bonding, stacking and solvent acting on the $j$th base pair. I will use the same parameter values $k=0.00045$\- eV/$A^2$ for the linear part of the base stacking interaction, $b=0.2 A^{-1}$ for the inverse range of the nonlinear base stacking, and $\alpha_{AT}=4.2 A^{-1}$,  $\alpha_{GC}=6.9 A^{-1}$ for the inverse ranges of the Morse potentials for all sequences discussed in this paper;  $\rho$, the relative strength of the nonlinear base stacking will be set equal to $50$, which is roughly comparable to the ratio of accepted values for the bending stiffness of double-stranded and single-stranded DNA\cite{elasticity}. This leaves only the depths of the Morse potential $D_{AT}$ and  $D_{GC}$ to be determined from the data (cf. below). Furthermore, I will assume that the chain is subject to free-end boundary conditions. 

Key thermodynamic quantities of interest are (i) the partition function
\begin{equation} 
Z_N = \int_{-\infty}^{\infty} dy_1 \cdots  dy_{N} e^{- H_P/(k_BT)}
\label{eq:Z}
\end{equation}
and (ii) the probability that the $r$th base pair is bound,
 \begin{eqnarray}
\nonumber
q_r &=&\frac{1}{Z_N} \int_{-\infty}^{\infty} dy_1 \cdots  \int_{-\infty}^{\infty} dy_{r-1}
 \int_{-\infty}^{y_c}dy_r  \\
 &&   
 \int_{-\infty}^{\infty}dy_{r+1} \cdots  \int_{-\infty}^{\infty} dy_{N}
e^{-H_P/(k_B T)}
\label{eq:probb} 
 \end{eqnarray}
where $k_B$ is the Boltzmann constant, $T$ the temperature and $y_c$ is an appropriately chosen displacement which distinguishes the open ($y>y_c$) from the closed  ($y<y_c$) state of a base pair. Unless otherwise stated, the choice will be $y_c= 2 A$.

As has been extensively discussed in the literature (e.g. \cite{Zhang})  the partition function (\ref{eq:Z}) diverges because of the flat top of the Morse potential; however, this divergence - ultimately responsible for the occurrence of an exact phase transition in the thermodynamic limit - is restricted to the disordered phase\cite{NTh2003}; as a result, 
 (\ref{eq:probb}) is always well defined - and can be shown to be numerically stable, i.e. independent of any upper cutoff of the integrations - for sufficiently long chains \cite{ShortChains}.

Numerical work was based on expanding \cite{Zhang, NTh2008} each factor in the partition function in terms of the complete set of eigenstates of the integral equation 
\begin{equation}
	\int_{-\infty}^{\infty} dy'K(y,y') \phi_{\nu}(y') = \Lambda_\nu \phi_{\nu}(y)
\label{eq:TI}
\end{equation}
with $K(y,y')=e^{-[W(y,y')+V_{AT}(y)/2+V_{AT}(y')/2]/k_BT}$ serving as a \lq\lq reference\rq\rq kernel. This transforms the multidimensional integrals in Eqs. (\ref{eq:Z}) and (\ref{eq:probb}) into matrix products in the reduced space of eigenstates of (\ref{eq:TI}) - typically of dimension 30-100 if one keeps only eigenstates with $\Lambda_{\nu}/\Lambda_{0}>10^{-8}$; the resulting matrix products can be calculated quite rapidly if one stores the intermediate results. 
Numerical computations have been performed by substituting the lower and upper limits of integration by $-1.5 A$ (owing to the rapid vanishing of the kernel which originates in the repulsive core of the Morse potential) and $300 A$, respectively; extending the region of integration in either direction does not produce any detectable change in the resulting melting profiles. Eq. \ref{eq:TI} has been discretized using Gauss-Legendre quadratures with a mesh of $1200$ points. Further numerical details will be reported elsewhere.
\section{Results}
\subsection{Melting profiles}
\begin{figure}[h]
\vskip -.75truecm
\resizebox{0.5\textwidth}{!}
{\includegraphics{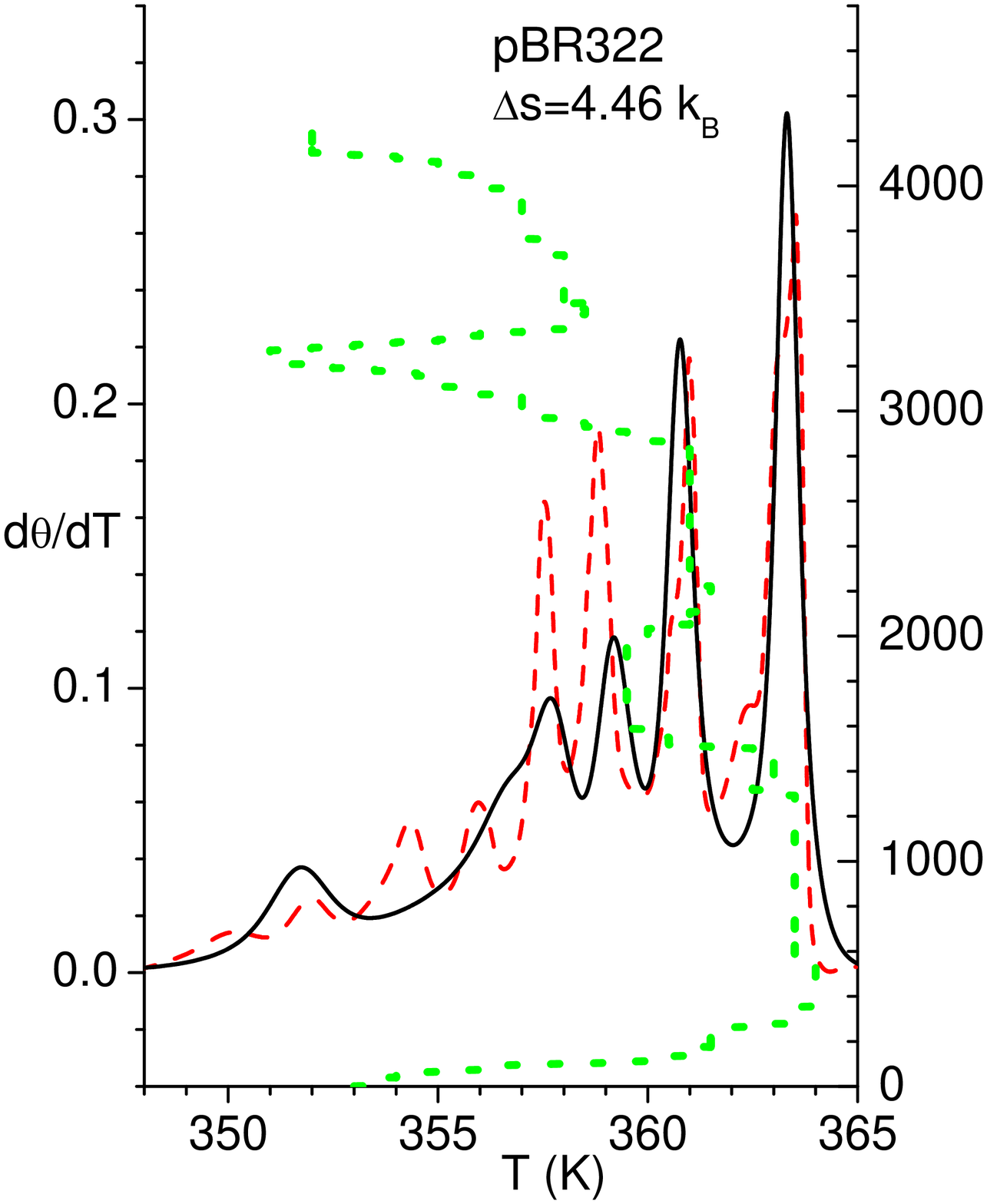}\includegraphics{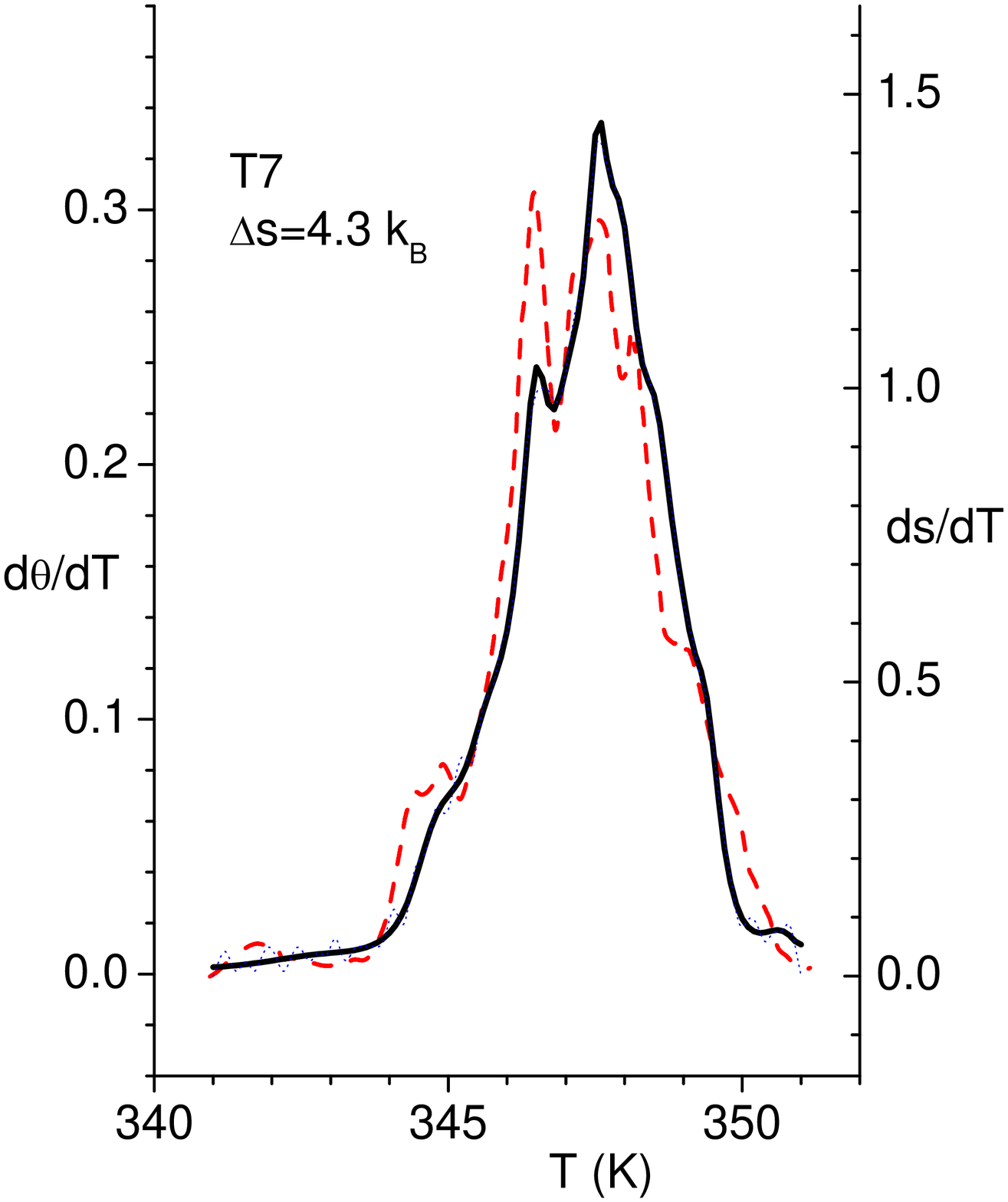}}	
\vskip -0.5truecm
\caption{ (color online)
{\small {\em Left panel}: left y-axis, melting curve for the plasmid  pBR322 (4361 bps), solid line PBD model calculation, dashed red line experimental data redrawn from  \cite{Blake91}; superimposed (dotted line) 
is the melting map, to be read as melting temperature vs. site (right vertical axis); 
{\em right panel}: computed melting profile (solid line, left y-axis) for the T7 phage (39937 bps); experimental data redrawn  from  \cite{FrankKame1976} (dashed red line); also shown 
(dotted curve, almost indistinguishable from the melting curve, right y-axis) is the temperature derivative of the entropy per site. 
}}
\label{fig:pBR322T7}
\end{figure}

The fraction of unbound ("melted") base pairs is given by
\begin{equation}
	\theta = 1 - \frac{1}{N}\sum_{i=1}^N q_i   \quad .
	\label{eq:meltfrac}
\end{equation}
Fig. \ref{fig:pBR322T7} shows calculated and experimental differential melting curves $d\theta /dT$ vs. $T$ for 
 (i) the plasmid pBR322 (4361 bps, 
GC content 53.8 \%, $\rm Na^+$ ion concentration $c_0=0.075 M$, 
$D_{AT}^0=0.1255$\- eV, $D_{GC}^0=0.1655$\- eV),  and 
(ii) the  the T7 phage (39937 bps, 48.4\% GC content, $\rm Na^+$ ion concentration $c=0.0195 M$, 
$D_{AT}=0.1205$\- eV, $D_{GC}=0.1619$\- eV).
The calculated melting curves, although not in perfect agreement with the experimental ones, reflect to a considerable extent the latter's complexities. Note in particular the multipeak structure on the left panel, and the correct overall shape of the melting curve on the right panel. The superimposed melting map shown in the left panel (melting temperatures, in steps of $0.5$ K, uniquely defined for each base pair via $p_i [T_m(i)]=1/2$) exhibits the distinct vertical regions  known from statistical (PS-type) theories of DNA denaturation to  coincide with peaks of the melting profile and to characterize cooperative melting of large domains, extending over hundreds of base pairs.

A careful analysis of the critical behavior of pure AT and GC sequences reveals essentially linear dependence of the melting temperature on $D$ in the range of interest. Since 
$T_m$ is known \cite{DelcourtBlake1998} to depend logarithmically on $c$,  
it is natural to assume that
\begin{equation}
\label{eq:DATpred}
D_{\sigma}  =   D_{\sigma}^0 + \kappa_{\sigma} \log (c/c_0) \quad, 
\end{equation} 
where $\sigma=AT,GC$, and use the two sets of Morse depth parametes to extract estimates of the proportionality constants, $\kappa_{AT}=0.00855\-$ eV and $\kappa_{GC}=0.00615$\- eV, respectively \cite{kappa}. This allows the calculation of $D_{AT}$ and $D_{GC}$ at any salt concentration, and therefore, in principle, the prediction of melting behavior for any long DNA sequence. 

\begin{figure}[h]
\vskip -.5truecm
\resizebox{0.5\textwidth}{!}
{\includegraphics{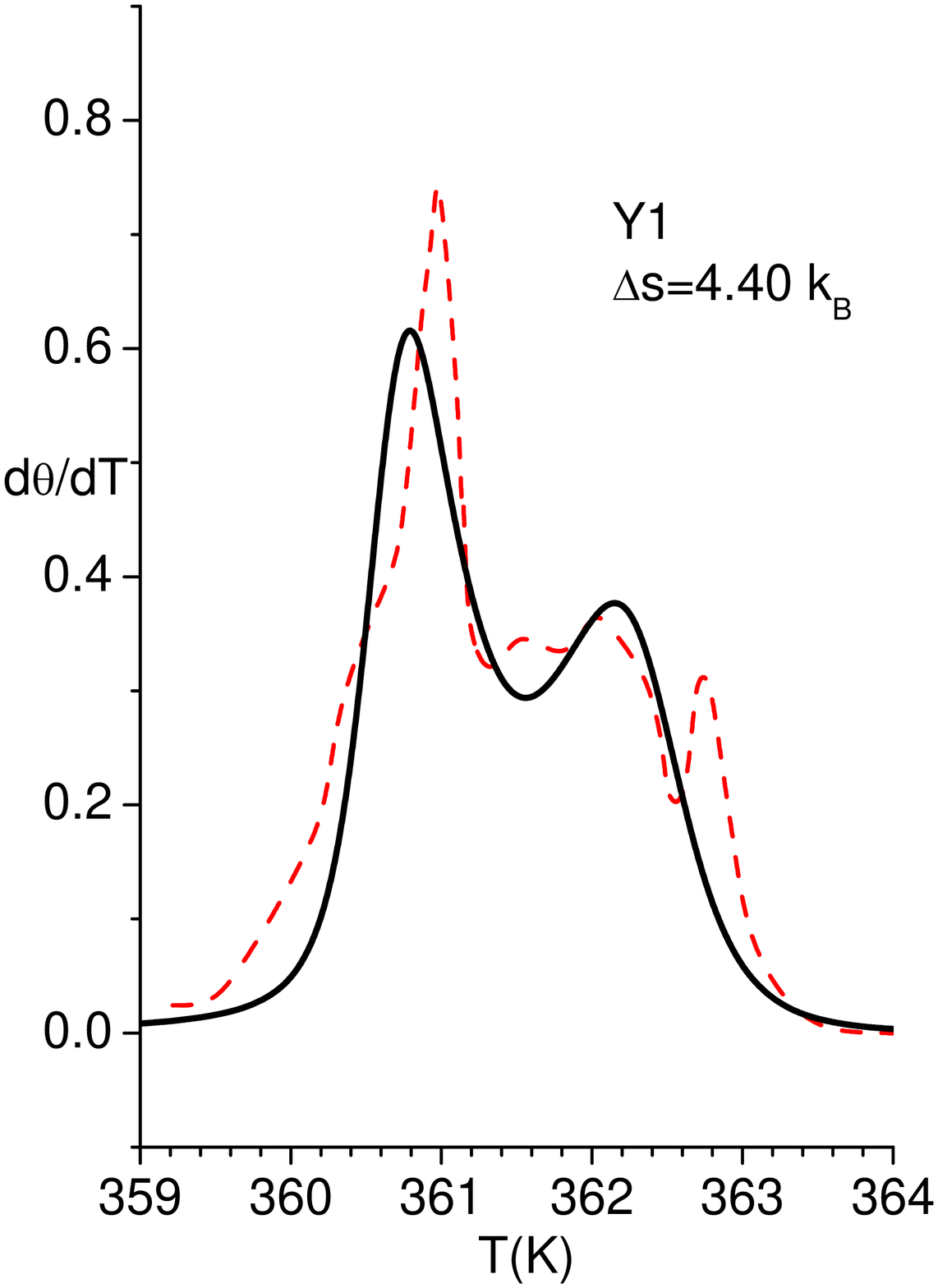}\includegraphics{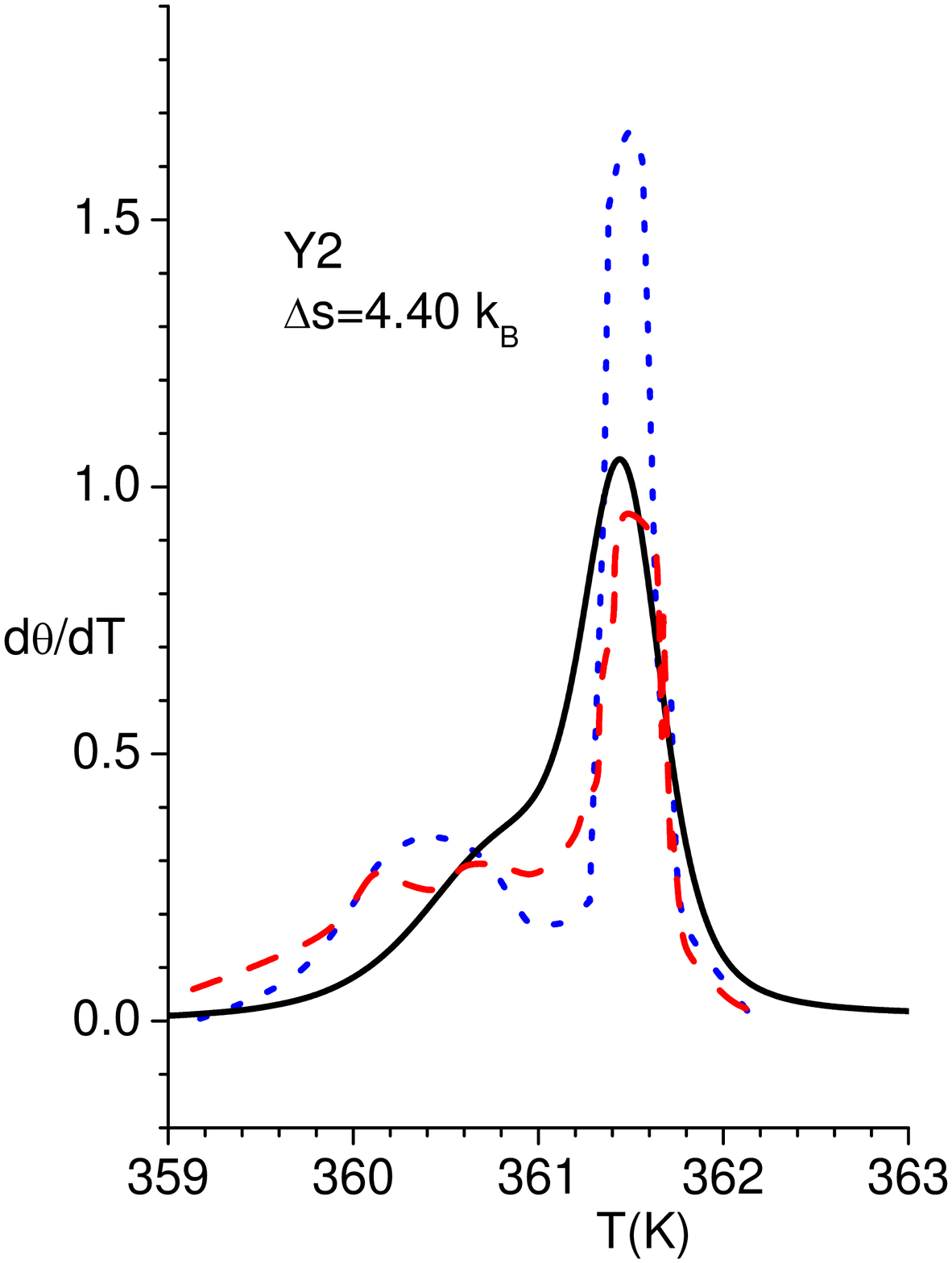}
}	
\vskip -0.5truecm
\caption{ (color online)
{\small {\it left panel}: Melting curves for Y1 fragment (2746 bps) of the $\phi$X174 phage, solid line PBD model prediction, 
dashed line experimental curve redrawn from \cite{WaBe85};
{\it right panel}: same for Y2 fragment (1695 bps), experimental data  redrawn from \cite{WaBe85} based on two different experiments (dashed line \cite{Perel81}, dotted line \cite{Tachi82}).  
}}
\label{fig:Y12}
\end{figure}

I have tested the validity of the above procedure in three cases.
Fig. \ref{fig:Y12} compares predicted to experimental melting profiles for the Y1 and Y2 fragments of the $\phi$X174 phage. The $\rm Na^+$ ion concentration is $c=0.195 M$ and the corresponding Morse depths, based on (\ref{eq:DATpred}), $D_{AT}=0.12905$ and $D_{GC}=0.16805$ respectively. Fig. \ref{fig:fdCarson} compares predicted to experimental melting profiles for the fd-phage (6408 bps, 40.9\% GC content). The $\rm Na^+$ ion concentration was \cite{Wada1976} $c=0.0195 M$, therefore the Morse depths used are the same those of the T7-phage (cf. above). In all three cases, PBD-based calculations with no adjustable parameters successfully predict melting temperatures and profiles in considerable detail. 
 
\subsection{Longer sequences}
\begin{figure}[h]
\vskip -.75truecm
\resizebox{0.5\textwidth}{!}
{\includegraphics{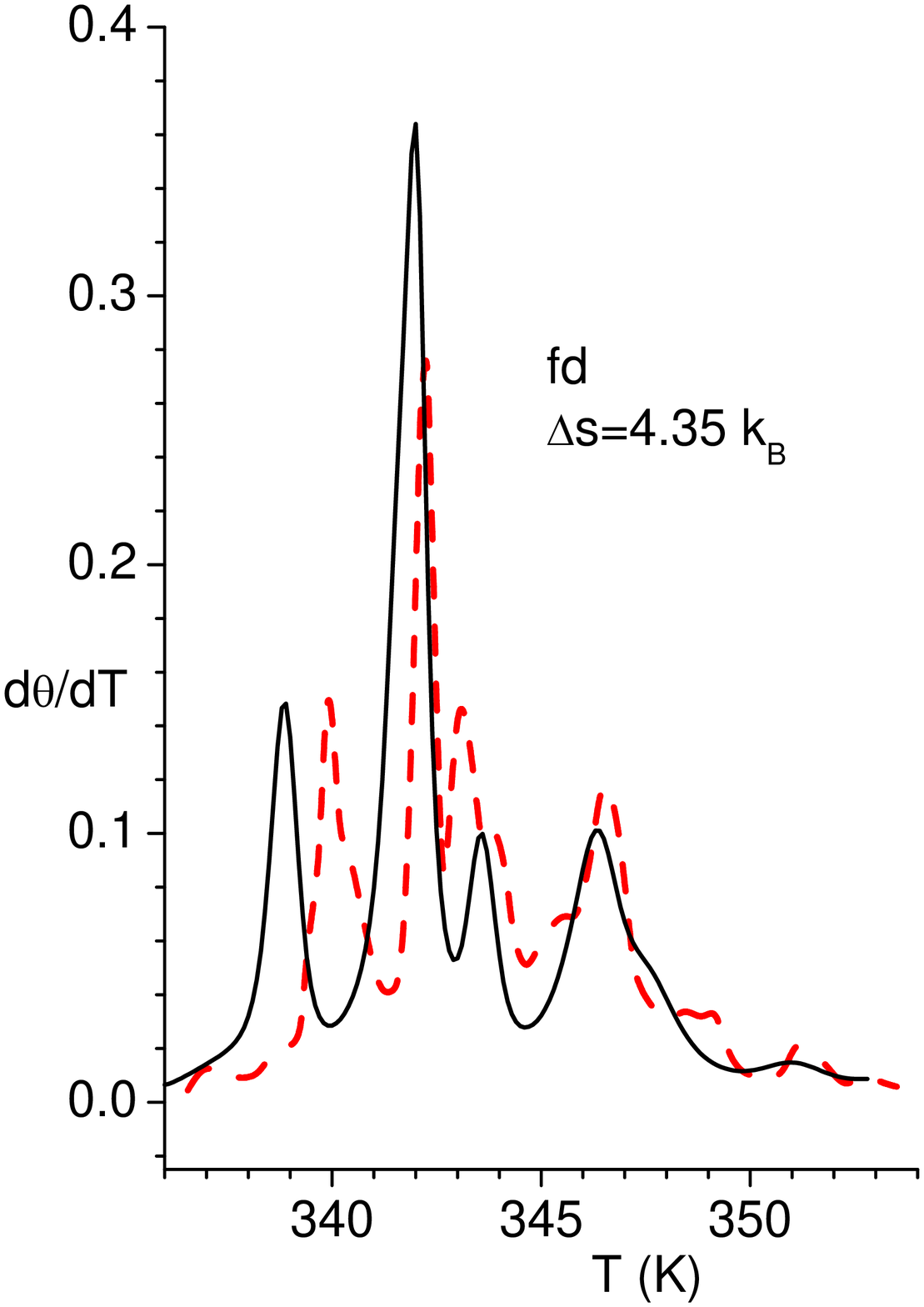}\includegraphics{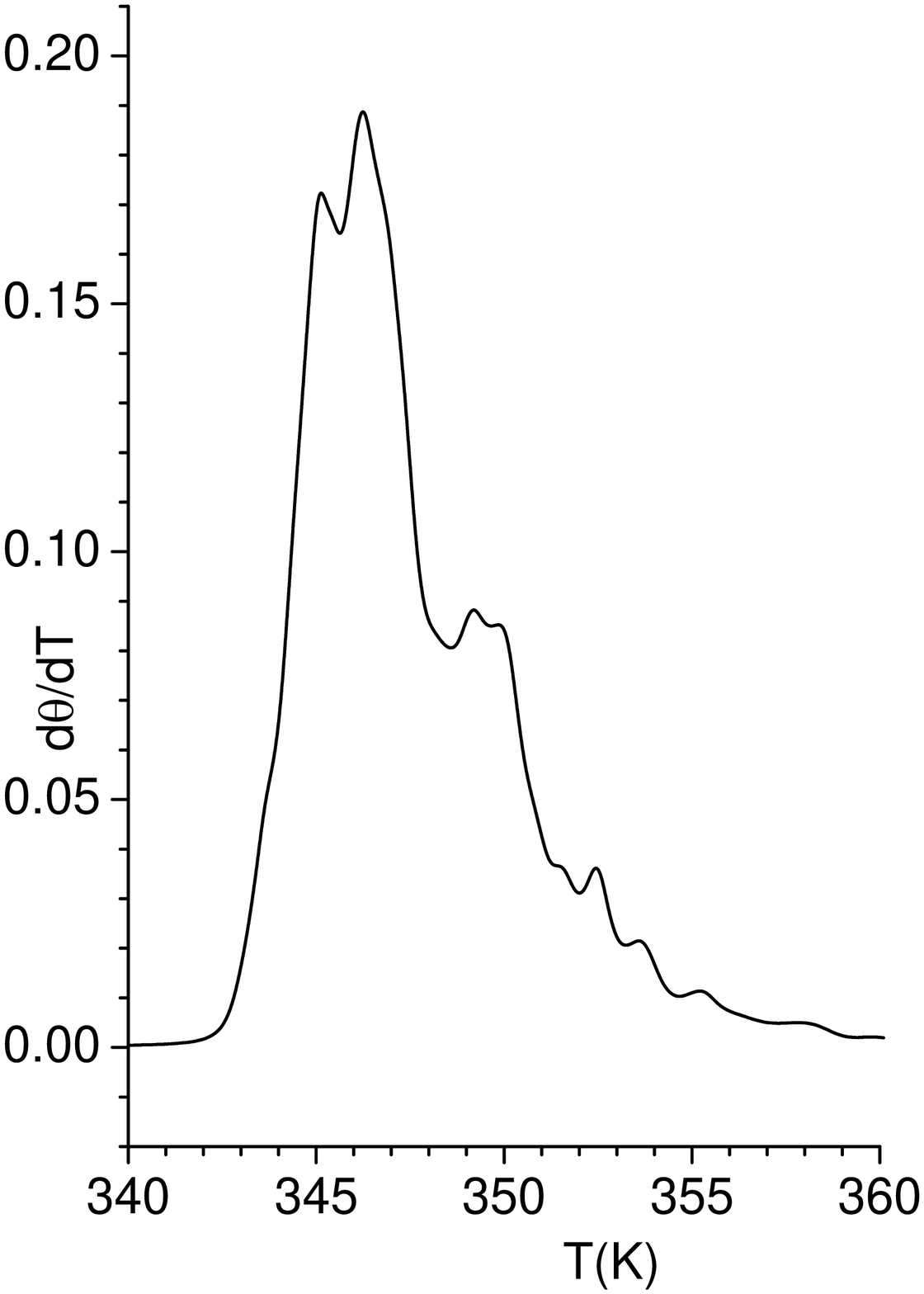}}
\vskip -0.5truecm
\caption{ (color online)
{\small  {\it Left panel:} Melting curve for the fd-phage (6408 bps), PBD model prediction (solid line) vs experiment (dashed line,  redrawn from\cite{Wada1976}). 
{\it Right panel:} Theoretical (PBD) melting profile of Carsonella Ruddii (159662 bps).
}}
\label{fig:fdCarson}
\end{figure}

The numerical method used here can in principle deal with much longer sequences. The right panel of Fig. \ref{fig:fdCarson} shows the theoretical melting profile obtained for the full genome sequence of the bacterial endosymbiont {\em  Carsonella ruddii} (159662 bps, GC content 16.6\%, shortest genome of any characterized bacteria  \cite{Carsonseq}) using - somewhat arbitrarily - the same parameters as in the Y1 and Y2 fragments (cf. above).  
The transition has a temperature width of almost $20$ K; this is unusually broad for such a large system and originates in large scale fluctuations of locally averaged GC-content. 

\subsection{Melting Entropy}
The entropic signature of the melting transition can be followed in some detail by looking at the temperature derivative $ds/dT$ of the entropy per site $s=[k_B /N] d(T\ln Z_N)/dT $. The right panel of Fig. \ref{fig:pBR322T7} shows that the proportionality relationship 
\begin{equation}
\frac{ds}{dT} = \Delta s \frac{d\theta}{dT}
\end{equation}
holds to a great degree of accuracy (the two curves being practically coincident), with $\Delta s$ being defined as the entropy of melting. Within the PBD model framework, calorimetry and base-pair sensitive methods (e.g. UV absorption) deliver essentially identical profile information. Numerical integration gives values $\Delta s=4.46, 4.35, 4.41 k_B, $ for the pBR322 plasmid, the fd and T7 phages respectively, and 4.40 $k_B$ for the Y1 and Y2 fragments; although considerably lower than typical experimental values \cite{Blake91} of 12\-$k_B$ (corresponding to 24 cal/mol(bp)/K), these values represent a significant improvement over previous estimates, around 1$k_B$,  obtained \cite{TDPprl} using oligomer-based  \cite{CampaGian} parameters. 
 
%

\subsection{Opening of individual base pairs}
\begin{figure}[h]
\vskip -.5truecm
\resizebox{0.3\textwidth}{!}
{
\includegraphics{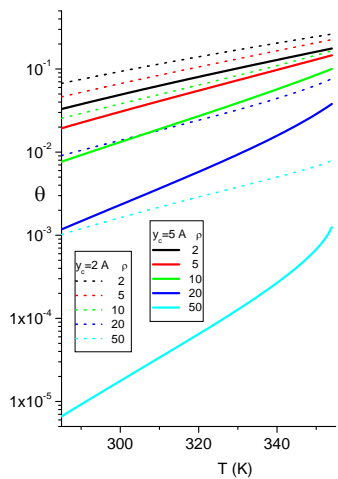}}
\vskip -0.5truecm
\caption{ (color online)
{\small 
Fraction of unbound base pairs as a function of temperatures calculated for the T7-phage
for $\rho=2, 5, 10, 20, 50$ and $y_c=2 A$ (dotted lines), $y_c=5 A$ (solid lines).  Only the lowest curve ($\rho=50, y_c=5 A$) leads to results consistent with estimates $\theta \sim 10^{-5}-10^{-6}$ obtained from imino proton exchange measurements at room temperature \cite{Gueron87}.
}}
\label{fig:Opening}
\end{figure}

Base pair opening has been carefully investigated \cite{Gueron87} by making use of the imino proton exchange technique. The experiments suggest (i) a closed base pair lifetime in the order of a few milliseconds and (ii) equilibrium constants between open and closed states corresponding to
a fraction of open base pairs of the order of 1 ppm (at room temperature). Since it is not immediately obvious what values of $y>y_c$ correspond to the open state in the imino proton experiment, I have plotted in 
Fig. \ref{fig:Opening} the melting fraction as a function of temperature for a range of $\rho$ values and two choices of critical amplitude $y_c$, equal to $2 A$ and $5 A$, respectively. The results indicate that the choice $\rho=50$, $y_c=5 A$ is consistent with the conclusions of the imino proton exchange experiments.  It should of course be noted that the exact choice of $y_c$ does not affect the overall shape or position of the melting profile (except for a sharpening of  the edges). Nonetheless, it is reassuring to confirm that a reasonable choice of $y_c$, when applied to the PDB model with the above (differential melting curve - based) parametrization leads to room temperature open base-pair populations which are in agreement with an entirely different class of experiments.  

\section{Concluding remarks}
In summary, the nonlinear lattice dynamics (PBD) approach to the statistical physics of DNA denaturation, in addition to elucidating the more abstract, qualitative aspects of the phase transition, has been shown to provide a powerful tool for the detailed description of quantitative features of melting in long genomic sequences. 

I acknowledge fruitful discussions and correspondence with Michel Peyrard.



\begin{thebibliography}{0}
\bibitem{kappa} It may be noted here that the inequality 
$\kappa_{AT} > \kappa_{GC}$ is in line with the well known property \cite{DelcourtBlake1998} that the spread $T_m(GC)-T_m(AT)$ decreases with increasing salt concentration.
\end{thebibliography}
\end{document}